# Encoding Propagation Invariance into Light

***Wenxiang Yan,[1,†] Tianyue Li,[2,†] Zhuolin Wu,[1] Yiyu Zhao,[1] Zhi-Cheng Ren,[1,3] Xi-Lin Wang,[1,3] Hui-Tian Wang,[1,3,4,*] Shuming Wang,[1,3,*] and Jianping Ding,[1,3,5,*]***

[1]National Laboratory of Solid State Microstructures and School of Physics, Nanjing University, Nanjing 210093, China

[2]Department of Physics and State Key Laboratory of Optical Quantum Materials, The Hong Kong University of Science and Technology, Clear Water Bay, Hong Kong SAR, China

[3]Collaborative Innovation Center of Advanced Microstructures, Nanjing University, Nanjing 210093, China

[4]Collaborative Innovation Center of Extreme Optics, Shanxi University, Taiyuan 030006, China

[5]Collaborative Innovation Center of Solid-State Lighting and Energy-Saving Electronics, Nanjing University, Nanjing 210093, China

[†]These authors contributed equally to this work.

*Corresponding author: htwang@nju.edu.cn; wangshuming@nju.edu.cn; jpding@nju.edu.cn;

## Abstract

Diffraction governs the axial evolution of optical fields, whereas conventional holographic synthesis primarily controls their transverse structure. Here we add axial diffraction management as an additional design freedom to the transverse-field programmability of holography, enabling the transverse optical function and its diffraction-driven axial evolution to be co-designed. By incorporating established propagation-invariant dynamics into computer-generated hologram and meta-hologram synthesis, we realize task-selectable axial responses in user-defined monochromatic, full-colour and vectorial fields. On a spatial light modulator, the same scalar user-defined field is configured either for a rapidly evolving 1.2-cm depth of field or for an approximately 75-cm propagation-invariant range. Millimetre-scale metasurfaces further enable metre-scale refocusing-free full-colour projection and vectorial colour fields with polarization textures preserved over more than 30 cm. This transverse–axial co-design framework extends holographic field synthesis beyond transverse programmability, providing a broadly compatible route towards task-configurable optical systems and compact multidimensional photonics.

## Introduction

Diffraction governs how optical fields evolve in space. In many optical applications, the transverse structure of a light field determines the optical function it carries, whereas its axial evolution determines where and over what range that function remains available. In optical systems, this evolution can blur images, redistribute optical energy and destabilize designed field structures when the observation plane, target or receiver is displaced (Supplementary Note 1). Many systems therefore use tunable lenses, mechanical actuation, adaptive optics or real-time tracking to maintain the required focal condition[1–5]. Although effective, these approaches can introduce additional components, calibration, power consumption and latency[1–6]. Rapid diffractive evolution is not universally undesirable, however. It is essential when axial selectivity and out-of-focus rejection are required, as in confocal microscopy[7], optical sectioning[8], and multilayer holographic displays[9]. The central challenge is therefore to manage diffractive evolution: to determine when a field should vary rapidly along the optical axis and when it should remain stable during propagation. This raises a broader question: can the desired axial diffractive evolution be encoded directly into the user-defined transverse light field?

Non-diffracting structured light[10–12] provides an important physical precedent. Bessel, Mathieu and Weber beams approximate propagation-invariant solutions of the Helmholtz equation over finite axial ranges. Their distinctive dynamics originate from conical wavevector distributions[10,11], corresponding to annular spatial-frequency spectra[13–16]. Such fields have enabled advanced optical trapping[17,18], extended-depth and self-healing imaging[19,20], high-aspect-ratio laser processing[21,22], and applications in nonlinear and atom optics[10,11]. However, propagation invariance has largely remained associated with specific analytical beam families or specialized field geometries. Their predefined transverse profiles limit function flexibility and can be accompanied by sidelobe-related trade-offs (Supplementary Note 2). A key opportunity is therefore to separate axial propagation invariance from these special beam profiles and make it available as a programmable property of user-defined optical fields.

Depth of field (DoF) provides a measurable manifestation of this axial diffractive evolution. It describes the range over which an optical field retains acceptable fidelity[23]. Different tasks require different diffractive responses. Short DoF provides axial discrimination for microscopy, optical sectioning and multilayer holography[7–9], whereas long DoF provides propagation robustness for defocus- or motion-tolerant displays, wearable optics, free-space communication, microscopy, projection, optical processing and metrology[19,24–28]. These regimes reveal a basic trade-off between axial selectivity and propagation robustness. Yet in many optical systems, the axial diffractive response is determined by the chosen hardware, optical configuration or beam family. Existing strategies provide powerful task-specific solutions through active refocusing, axial scanning, fixed extended-DoF optics, computational recovery or specialized structured-light fields[2,3,8,9,29–38], but they generally operate within predetermined axial regimes (Supplementary Note 3). Switching between fundamentally different axial functions—such as rapid axial evolution for selectivity and extended propagation invariance for robustness—commonly requires a corresponding change in the optical configuration, control strategy or field construction. The missing capability is therefore a general field-synthesis framework that co-designs a user-defined transverse structure with the desired axial diffractive evolution, allowing the same field to be configured for rapid axial evolution or extended propagation invariance according to the task, without optical-train modification.

Here we address this capability by incorporating established propagation-invariant dynamics into computer-generated hologram (CGH) synthesis, adding axial diffraction management to its transverse-field programmability. The hologram spatial-frequency bandwidth then controls the axial evolution of the reconstructed transverse field: broad spatial-frequency support produces the rapidly dephasing, short-DoF regime, whereas narrower annular bandwidth progressively extends the propagation-invariant range (Figs. 1A–1D). We first demonstrate this principle for monochromatic scalar fields using a spatial light modulator. Without additional refocusing hardware or optical-train modification, the same target field can be configured from a conventional short DoF of 1.2 cm to a propagation-invariant range of 75 cm, corresponding to a more than 60-fold extension (Fig. 2). Thus, our framework provides a means to co-design a user-defined transverse field and its task-selectable axial evolution within the same field-generation architecture.

We further extend this framework from scalar fields to multidimensional light. Practical optical systems often require coordinated axial diffractive evolution across multiple wavelengths, polarization channels and vectorial field structures. These components generally evolve differently under diffraction, making a shared propagation-invariant evolution difficult to realize within a compact cascade system. We therefore encode the same principle into wavelength- and polarization-multiplexed metasurfaces. Within millimetre-scale planar devices, different spectral and polarization channels are assigned coordinated axial diffractive evolution, enabling metre-scale refocusing-free full-colour projection (Fig. 3) and defocus-tolerant vectorial colour fields that preserve their polarization textures over more than 30 cm (Fig. 4). Conceptually, this work transfers propagation-invariant dynamics from a limited set of analytical beam profiles to user-defined monochromatic, polychromatic and vectorial transverse fields. Technically, it shifts control of the axial diffractive response from external compensation and optical-system configuration to field-encoded transverse–axial co-design. Functionally, it allows the same field-generation platform to be configured for axial selectivity or propagation robustness according to different optical tasks. This capability provides a broadly compatible route towards task-configurable optical systems and compact multidimensional photonic platforms for display, imaging, projection, optical manipulation, processing, free-space communication and information encoding[1–6,8,19,20,27,28,39–49].

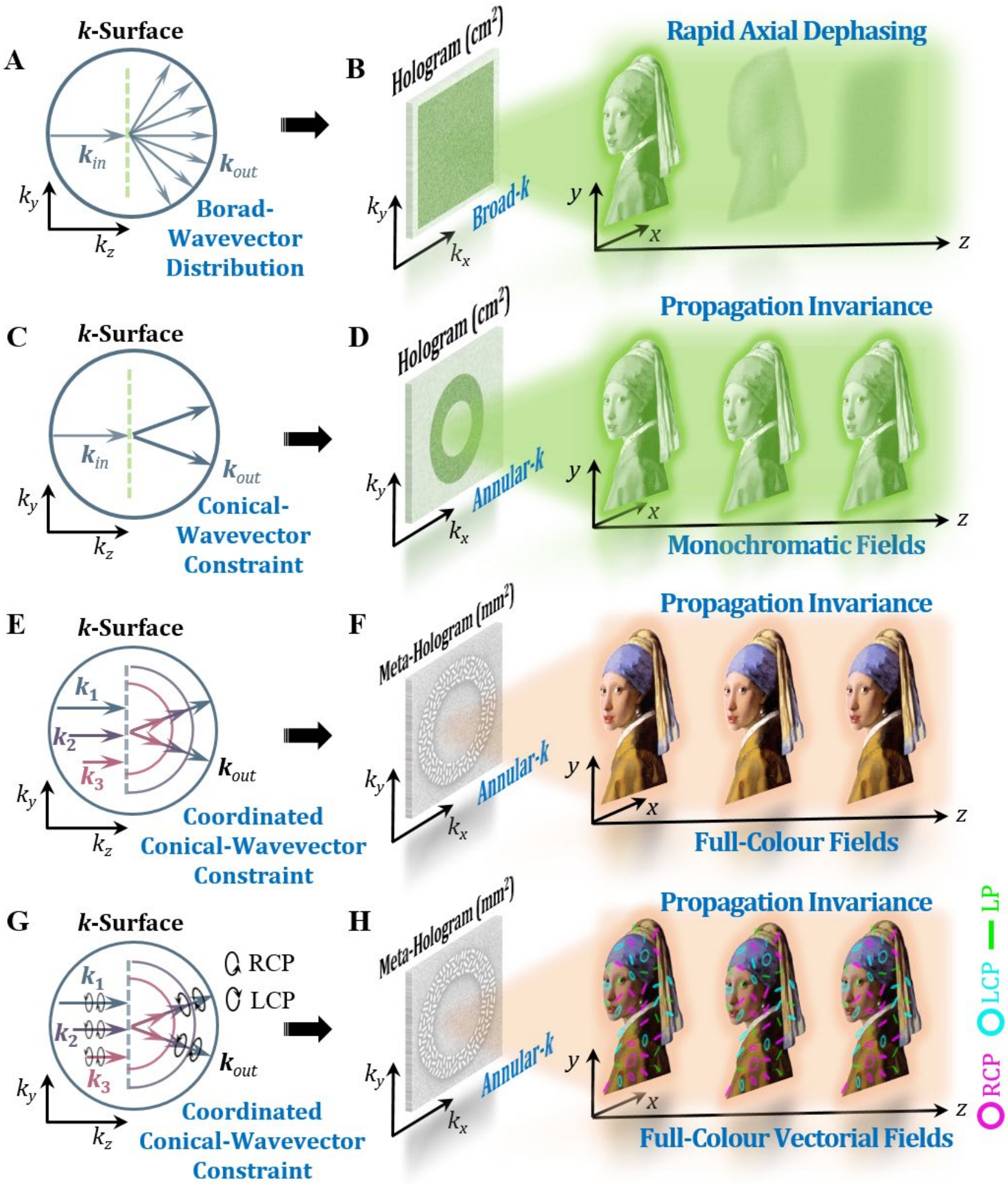

**Fig. 1 | Wavevector-domain encoding of propagation invariance. A–B,** Conventional computer-generated holography (CGH) uses default broad spatial-frequency support (Broad-***k***). The output wavevectors have different longitudinal components $k_z$ on the ***k***-surface, or Ewald sphere (**A**), producing rapid axial dephasing when the observation plane moves away from the designed reconstruction plane (**B**). $\boldsymbol{k}_{in}$ and $\boldsymbol{k}_{out}$, input and output wavevectors of monochromatic light. **C–D,** Encoding propagation invariance in a monochromatic field. Constraining the output wavevectors to a conical distribution on the ***k***-surface (**C**) corresponds to an annular spatial-frequency band (Annular-***k***) in the Fourier-domain hologram (**D**), preserving the interference condition over a selectable axial range. **E–F,** Extension to full-colour fields. Coordinated conical-wavevector constraints are assigned to different wavelengths within a single millimetre-scale annular meta-hologram, producing a shared propagation-invariant response across spectral channels. **G–H,** Extension to colorful vectorial fields. Wavelength and spin channels are jointly coordinated within one meta-hologram, enabling propagation-invariant reconstruction of multidimensional optical fields. $\boldsymbol{k}_1$, $\boldsymbol{k}_2$, and $\boldsymbol{k}_3$, input wavevectors at different wavelengths. RCP, right-handed circular polarization; LCP, left-handed circular polarization; LP, linear polarization.

# Results

## Wavevector-Domain Encoding of Propagation Invariance

In Fourier optics, an optical field can be decomposed into plane-wave components with different transverse and longitudinal wavenumbers. During propagation, differences in the longitudinal wavenumbers produce relative phase accumulation among these components, thereby governing the diffractive evolution of the field. Conventional computer-generated holography (CGH) uses broad spatial-frequency support for flexible transverse-field synthesis. However, the broad angular spectrum also leads to rapid axial dephasing when the observation plane moves away from the designed reconstruction plane[1–6,8,9] (Figs. 1A and 1B). By contrast, propagation-invariant structured light uses conical wavevector distributions, or narrow annular spectra, so that the constituent plane waves have closely matched longitudinal wavenumbers and preserve their interference condition over an extended axial range. This provides intrinsic propagation robustness, although the resulting transverse profiles are usually associated with specific analytical beam families, such as Bessel modes, and accompanied by sidelobes[10,11,13–16] (Supplementary Note 2). Conventional CGH and propagation-invariant structured light therefore represent two complementary capabilities—transverse-field programmability and axial propagation invariance, respectively (Supplementary Note 4 and Table S2). Together, these capabilities correspond to two essential aspects of optical-field function: the transverse structure determines what the field does, whereas the axial evolution determines where and over what range that function remains available.

Our method brings these complementary capabilities into a common synthesis framework by incorporating the established conical-wavevector condition into CGH synthesis, making axial propagation invariance designable for user-defined transverse fields (Figs. 1C and 1D). In the Fourier domain, the holographic spectrum is constrained within an annular spatial-frequency band. The bandwidth of this annulus determines the spread of longitudinal wavenumbers and therefore the rate of diffractive evolution: broader bandwidth recovers the rapidly dephasing regime of conventional CGH, whereas narrower bandwidth progressively extends the propagation-invariant range. Importantly, the broad-bandwidth regime is not merely an inferior reference: its rapid axial evolution is functional when axial discrimination or plane-selective reconstruction is required. This bandwidth control therefore adds axial diffraction management as an additional degree of freedom to the transverse-field programmability of CGH, allowing the transverse optical function and its axial evolution to be co-designed within the same synthesis architecture. The annular bandwidth thereby acts as a hologram-encoded control parameter for axial diffraction, shifting the determination of the propagation response from optical-system configuration to field synthesis itself[2,3,29,30,33,36–38]. (Methods, Supplementary Notes 3 and 4 and Tables S1 and S2)

We first validate this principle in monochromatic scalar fields using a phase-only spatial light modulator (SLM). For a standard yin–yang pattern, conventional CGH with broad spatial-frequency support reconstructs the target field near the designed plane but degrades rapidly during propagation, corresponding to a measured DoF of 1.2 cm (Figs. 2A–2D). When the conical-wavevector condition is incorporated into hologram synthesis, the same target profile remains propagation-invariant over approximately 75 cm without optical-train modification (Figs. 2E–2H and Supplementary Movie 1). This represents a more than 60-fold extension of the usable axial range. Here, DoF provides the quantitative measure of propagation invariance and is defined as the axial range over which the Pearson correlation coefficient (PCC) between the reconstructed and target intensity fields remains above 90% of its peak value[23] (see Methods, "Quantification of depth of field").

Because axial diffractive evolution is controlled in the wavevector domain, multiple conical bands can also be integrated within a single kinoform. We demonstrate this by decomposing a composite image into two portraits—Dennis Gabor and *Girl with a Pearl Earring*—and assigning them to distinct annular bands with different longitudinal wavenumbers. The two portraits are encoded in separate spatial-frequency regions of one hologram and reconstructed simultaneously while maintaining propagation invariance over approximately 54 cm (Figs. 2I–2L). This multi-cone implementation improves spatial-frequency utilization while retaining the designed axial propagation behaviour. These SLM experiments establish the scalar proof of principle: propagation invariance represents one accessible axial regime, while more broadly, the transverse optical function and its diffraction-driven axial evolution can be co-designed through hologram-bandwidth control. We next extend this combined transverse–axial design capability to wavelength, polarization and vectorial degrees of freedom using metasurfaces.

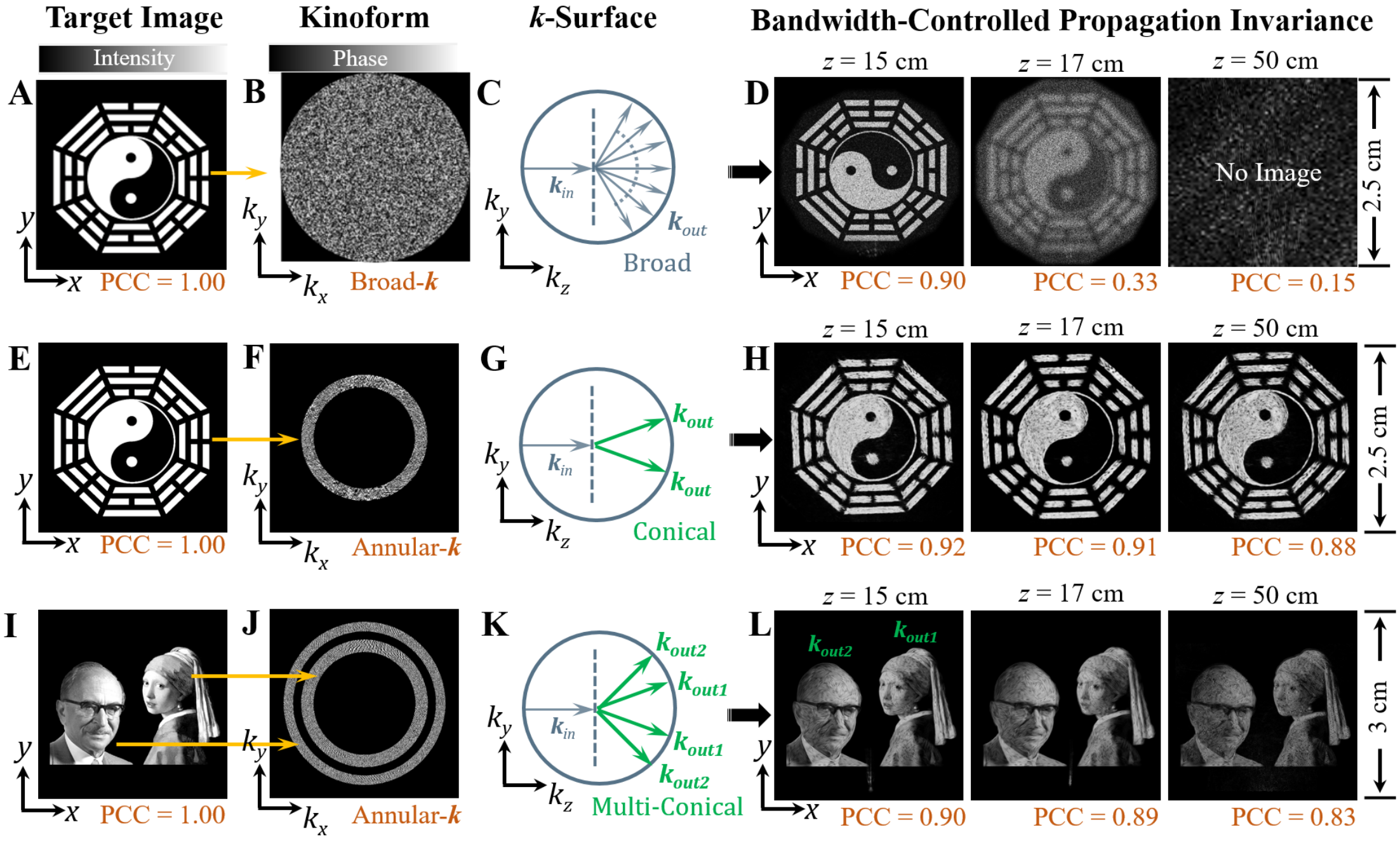

**Fig. 2 | SLM validation of bandwidth-encoded propagation invariance. A–D,** Conventional Fourier CGH[23] of a yin–yang target. **A,** Target image. **B,** Phase-only Fourier-domain kinoform. **C,** Broad output-wavevector distribution on the $k_y$-$k_z$ section of the $\boldsymbol{k}$-surface. **D,** Experimental reconstructions at $z$ =15 cm, 17 cm and 50 cm, showing rapid axial degradation and a measured DoF of 1.2 cm. **E–H,** Reconstruction of the same target using an annular spatial-frequency band. The output wavevectors are constrained to a conical distribution, allowing the reconstructed field to remain propagation-invariant over a measured DoF

of approximately 75 cm. **I–L,** Multi-cone encoding. A composite image containing Dennis Gabor and *Girl with a Pearl Earring* is decomposed and assigned to two annular bands corresponding to distinct conical distributions with different wavevectors $\boldsymbol{k}_{out1}$ and $\boldsymbol{k}_{out2}$. The two portraits are reconstructed from a single kinoform while maintaining propagation invariance over approximately 54 cm. The focal plane is $z$ = 15 cm. The Pearson correlation coefficient (PCC) quantifies axial fidelity, and DoF is defined as the axial range over which the PCC remains above 90% of its peak value[23]. See Supplementary Note 7 and Supplementary Movie 1 for details.

## Metasurface Encoding of Propagation Invariance in Multidimensional Fields

The SLM experiments establish wavevector-domain encoding of propagation invariance in monochromatic scalar fields. Practical optical systems, however, often combine multiple wavelengths, polarization channels and vectorial field structures. Because these components accumulate different propagation phases and exhibit different diffractive evolution[1–6,39–48,50], encoding a coordinated axial evolution across them is considerably more demanding than for a scalar field. Metasurfaces provide a compact platform for this purpose. Their subwavelength meta-atoms can tailor phase, amplitude and polarization responses within an ultrathin planar device, while multiplexing distinct optical functions across wavelength and polarization channels[25,39–48,51–65]. We therefore transfer the same propagation-invariance design principle to meta-holograms. By assigning coordinated conical-wavevector constraints to different colour and polarization channels, the desired axial response of user-defined multidimensional transverse fields is encoded directly into a single planar device (Figs. 1E–1H). This implementation is not merely a change of platform; it demonstrates that the transverse optical function and axial propagation response can be co-designed and coordinated across wavelength, polarization and vectorial degrees of freedom within an ultrathin architecture.

**Refocusing-free full-colour projection with metre-scale propagation invariance.** We first demonstrate this concept in a refocusing-free full-colour meta-projector. The target colorful image is decomposed into red, green and blue channels at working wavelength of 638, 561 and 488 nm, respectively. Each channel is rescaled to pre-compensate chromatic aberration[46,66,67] and assigned an annular phase profile that encodes the required propagation-invariant response. The three colour-dependent phase profiles are then integrated into a single dielectric meta-hologram composed of rectangular silicon nitride meta-atoms (Figs. 3A–3C). Distinct polarization-conversion channels—RCP→LCP, LCP→RCP and RCP→RCP—are used for the three colours to reduce inter-channel crosstalk and preserve colour fidelity[46] (see Supplementary Note 8). The resulting meta-hologram coordinates the axial evolution of the RGB components within one millimetre-level planar device, allowing the reconstructed full-colour field to remain stable over a metre-scale propagation range (Figs. 1E and 1F).

To increase information capacity while preserving propagation invariance, we generalize the design from a single conical band to multiple independent bands. The metasurface is divided into eight concentric annular zones, each corresponding to a distinct longitudinal wavenumber and propagation-invariant range (Fig. 3D). Within each zone, the RGB components are coordinated to the designated conical band, allowing one full-colour image to be encoded in each annulus. A single meta-hologram therefore projects eight independent full-colour images from the "Dynamics of a Fall" sequence, all maintaining high fidelity over a metre-scale axial range (Figs. 3F–3H and Supplementary Movie 2). This multi-cone implementation shows that propagation invariance can be multiplexed across both spectral and spatial-frequency channels, while improving spatial-frequency utilization[1–4,6,46,66,67] (see Supplementary Note 9).

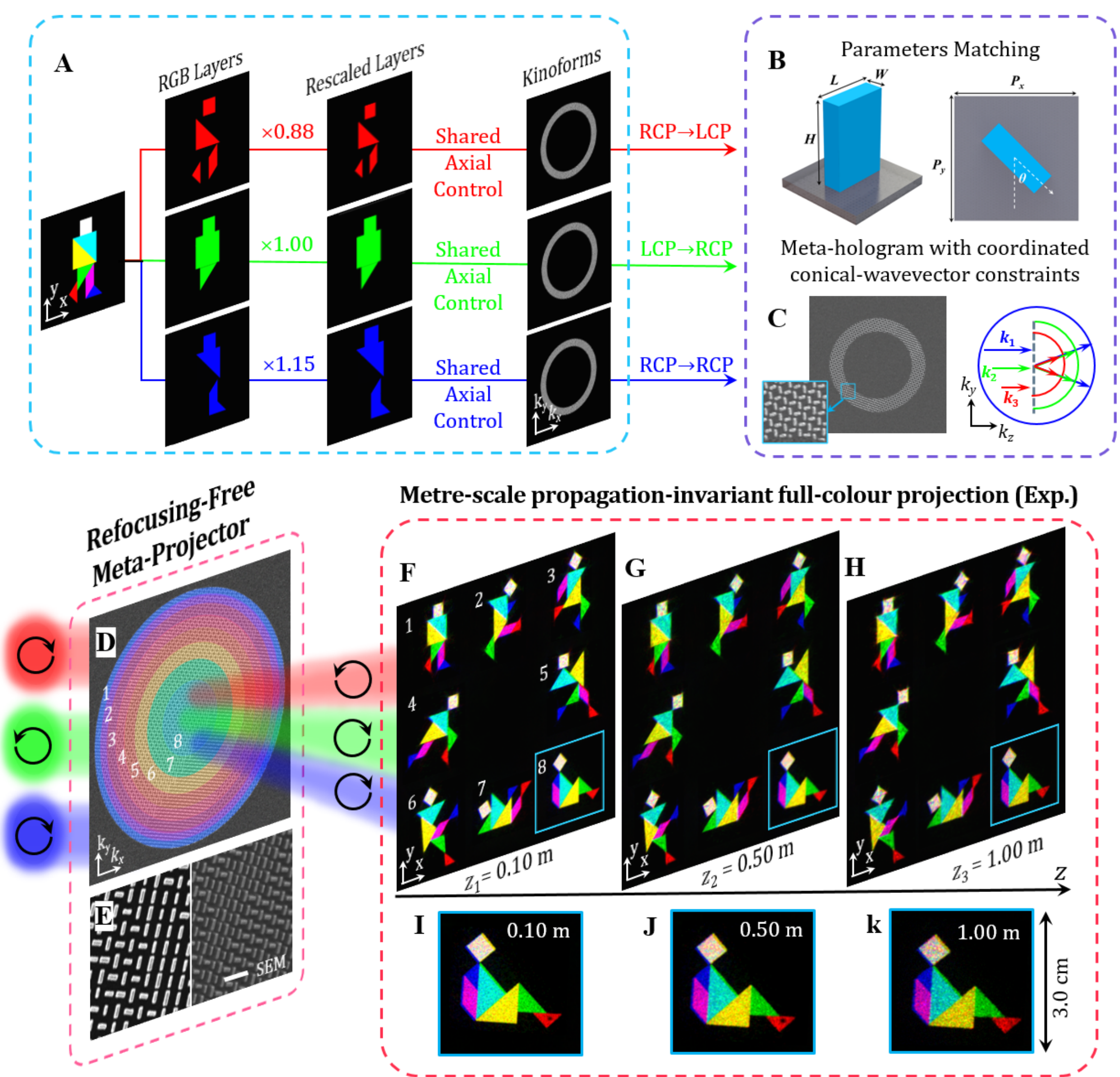

**Fig. 3 | Metre-scale propagation-invariant full-colour projection. A,** Design workflow for full-colour meta-hologram with coordinated axial propagation across red, green and blue channels. Wavelength-dependent scaling factors of 0.88, 1.00 and 1.15 are applied to the red, green and blue layers, respectively, to pre-compensate chromatic dispersion. **B,** Rectangular silicon nitride meta-atom with geometric parameters $L$, $W$ and $\theta$, fixed height $H$, and a 500-nm pixel pitch ($P_x$ = $P_y$ =500nm). **C,** Annular meta-hologram in which different wavelengths are assigned coordinated conical-wavevector constraints, encoding a shared propagation-invariant response. **D,** Layout of the 3 × 3 mm$^2$ meta-projector, divided into eight concentric annular zones. Each zone corresponds to a distinct conical band and encodes one full-colour image. **E,**

Scanning electron microscopy image of the fabricated metasurface. Scale bar, 1 μm. **F–H,** Experimental refocusing-free projection of eight full-colour images from the "Dynamics of a Fall" sequence: walking, quick walking, jogging, running, losing balance, slipping, falling down and sitting down. The colour–polarization channels are red, RCP→LCP; green, LCP→RCP; and blue, RCP→RCP. **I–K,** Enlarged experimental reconstructions of the "Sitting Down" image at different propagation distances, demonstrating metre-scale axial stability. Exp., experimental result. See Supplementary Note 9 and Supplementary Movie 2 for details.

**Propagation-invariant vectorial colour fields.** We next extend the metasurface implementation to simultaneous control over colour and polarization. In this design, colour multiplexing and polarization control are decoupled. The full-colour target is first separated into RGB layers with chromatic pre-compensation[46,66,67]. For each layer, an annular phase profile is generated, corrected for grating dispersion (Supplementary Note 10), and combined through complex-amplitude modulation[68] into a kinoform that reconstructs the full-colour field in one polarization-conversion channel[66,67,69], such as RCP→LCP. A second orthogonal channel, such as LCP→RCP, is encoded into the same meta-atoms with an independent kinoform. The two spin channels therefore carry independent RGB fields while sharing coordinated conical-wavevector constraints, producing a propagation-invariant vectorial colour field from a single-layer planar meta-hologram (Figs. 1G and 1H).

To encode multiple vectorial colour patterns, we fabricate a 1.2 × 1.2 $mm^2$ metasurface divided into seven concentric annular regions (Fig. 4A). Each annulus corresponds to a distinct conical band and encodes a customized vectorial colour pattern through two orthogonal polarization channels. Under linearly polarized white-light illumination, the reconstructed field in the first diffraction order preserves high-fidelity colour and complex polarization textures over an axial range exceeding 30 cm. We demonstrate two representative schemes: tailored vectorial distributions (Figs. 4B–4D and 4H–4J) and vectorial colour fields spanning the full Poincaré sphere of polarization states (Figs. 4E–4G and 4K–4M). Both schemes preserve their designed polarization textures during propagation, demonstrating propagation-invariant and defocus-tolerant vectorial reconstruction at visible wavelengths (see Supplementary Note 11 and Supplementary Movie 3).

Together, these metasurface demonstrations generalize wavevector-domain propagation-invariance encoding from programmable scalar holographic fields to multidimensional light. By coordinating the axial diffraction management of wavelength and polarization channels, millimetre-scale planar devices realize metre-scale refocusing-free full-colour projection and propagation-invariant vectorial colour fields over more than 30 cm. The axial robustness is encoded directly into the meta-hologram rather than maintained through external refocusing or optical-train modification (Supplementary Note 3). Metasurface integration therefore provides a compact route for co-designing user-defined transverse optical functions and coordinated axial propagation responses in polychromatic and vectorial fields, thereby extending multidimensional planar photonics from transverse-field programmability to joint transverse–axial field design[1–6,39–48].

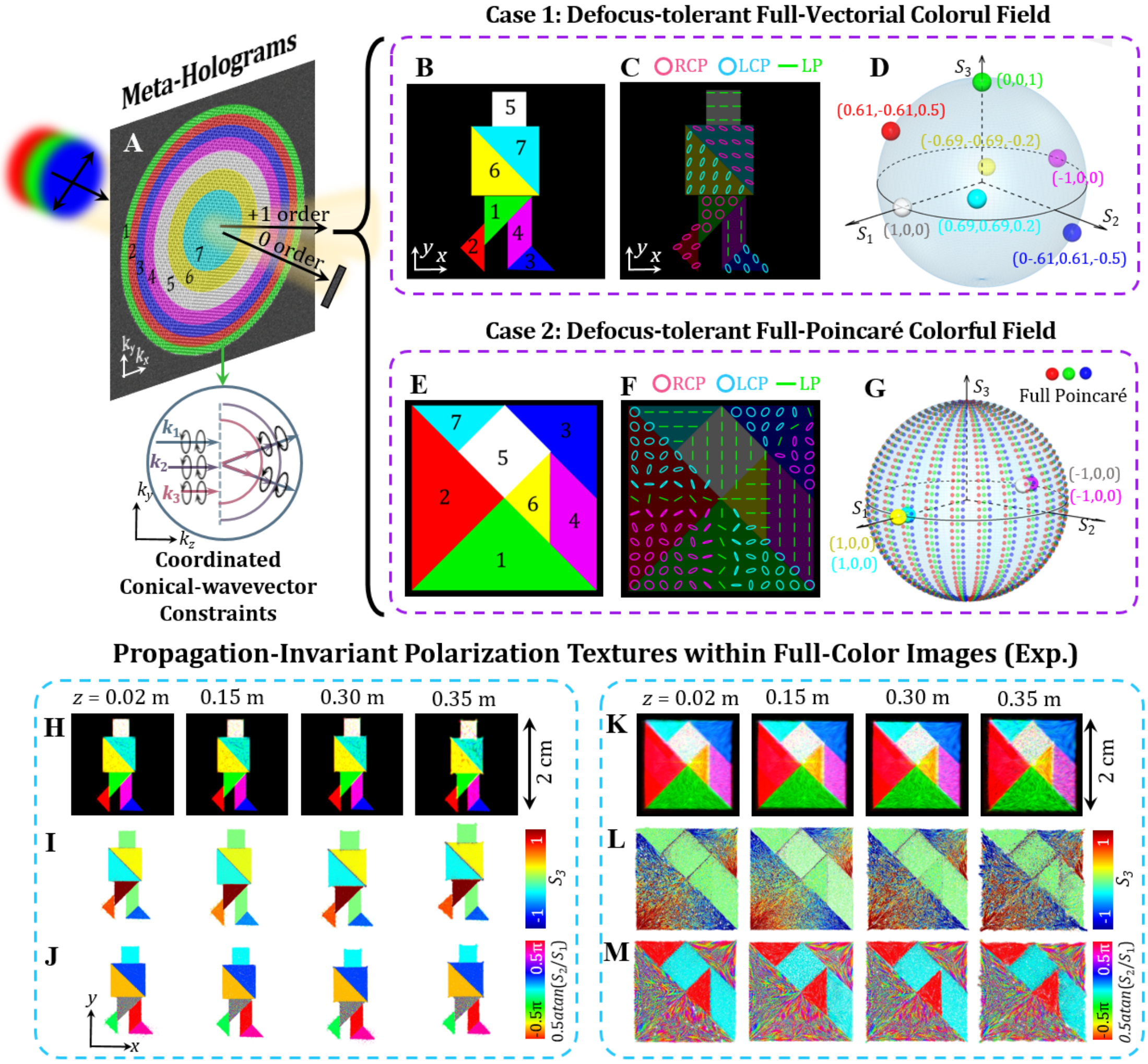

**Fig. 4 | Propagation-invariant vectorial colour fields. A,** Schematic of a 1.2 × 1.2 mm$^2$ meta-hologram divided into seven concentric annuli. Each annulus corresponds to a distinct conical band shared by three wavelengths and two orthogonal spin channels, and encodes one vectorial colour segment. $\boldsymbol{k}_1$, $\boldsymbol{k}_2$, and $\boldsymbol{k}_3$, input wavevectors at different wavelengths. **B–D,** Case 1: design of tailored full-vectorial colour fields. **B,** Target colour images. **C,** Target vectorial distributions. **D,** Corresponding Stokes parameters mapped onto the Poincaré sphere. **E–G,** Case 2: design of full-Poincaré vectorial colour fields. **E,** Target colour images. **F,** Target vectorial distributions. **G,** Stokes parameters showing coverage of polarization states on the Poincaré sphere for segments 1–3. **H–J,** Experimental results for Case 1, showing propagation-stable colour images (**H**), measured $S_3$ distributions (**I**) and polarization-orientation [0.5·atan($S_2$, $S_1$)] maps (**J**) over an axial range of approximately 30 cm. **K–M,** Corresponding experimental results for Case 2, demonstrating preservation of the designed colour and polarization textures during propagation. Exp., experimental result. See Supplementary Note 11 and Supplementary Movie 3 for details.

## Discussion

Diffraction governs the axial evolution of light, but in most optical systems the resulting response is determined indirectly by the hardware, optical configuration or beam family. More generally, the transverse structure of a field determines the optical function it carries, whereas its axial evolution determines where and over what range that function remains available. Here we bring these two aspects into a common field-synthesis framework. By incorporating established propagation-invariant dynamics into hologram and meta-hologram synthesis, the desired axial response—from rapid evolution to propagation invariance—is encoded into user-defined transverse fields. The annular spatial-frequency bandwidth controls the longitudinal-wavenumber spread: broader bandwidth produces a rapidly dephasing, short-range response, whereas narrower bandwidth progressively extends the propagation-invariant range. The central advance is therefore not simply extended DoF, but the addition of axial diffraction management as a field-encoded design freedom alongside transverse-field programmability, enabling the transverse optical function and its diffraction-driven axial evolution to be co-designed for the task. Experimentally, this principle spans from a 1.2-cm rapidly evolving reconstruction to a 75-cm propagation-invariant range on an SLM, and extends to metre-scale refocusing-free full-colour projection and propagation-invariant vectorial colour fields over more than 30 cm in millimetre-scale meta-devices.

This capability is important because rapid axial evolution and propagation invariance serve complementary optical functions. The former provides axial selectivity, depth discrimination and out-of-focus rejection, as required in confocal microscopy[7], optical sectioning[8], and multilayer holographic displays[9]; the latter provides extended working range, defocus tolerance and propagation robustness, as required in projection, metrology, imaging, laser processing, motion-tolerant displays, and free-space communication[1–6,19,24,25,27,28]. The objective is therefore not to favour either regime universally, but to select the axial response appropriate to the task while retaining the user-defined transverse optical function. In programmable systems, this shifts task adaptation from optical-train reconfiguration to field synthesis. This task-oriented use of holographic degrees of freedom is conceptually reminiscent of Benton holography, in which vertical parallax is deliberately removed to enable white-light viewing[49] (Supplementary Note 3). Here, however, the controlled quantity is the longitudinal-wavenumber spread, which is assigned during hologram synthesis to prescribe axial diffractive evolution while retaining the complete transverse field.

This task-selectable transverse–axial design also suggests concrete system-level opportunities. In free-space optical communication, diffraction-driven field spreading imposes a fundamental trade-off among propagation distance, beam divergence and receiver aperture[27,28]. By allowing propagation robustness to be encoded into user-defined high-dimensional information-carrying fields, our framework could relax this trade-off and potentially broaden the design space for robust structured-light links[70–72]. In near-eye displays, AR/VR and head-up displays, diffraction-induced defocus blur makes image quality sensitive to axial displacement and can require additional focal management, creating a tension with system compactness[1–6] (Supplementary Note 1). Encoding a propagation-invariant response directly into the holographic field offers a field-level route towards refocusing-free, motion-robust display functionality. Similar opportunities arise in imaging[7,8,24], optical manipulation[17,18,73] and laser processing[21,22], where transverse structure defines the optical function and axial evolution defines the usable depth or interaction range.

The framework also changes how non-diffracting structured-light physics can be used. Classical Bessel, Mathieu and Weber beams exhibit long-range propagation, self-healing and axial robustness[10–12,74,75], but these properties are tied to particular analytical field families and can be accompanied by characteristic sidelobes that reduce imaging contrast or produce unwanted processing effects[76–81] (Supplementary Note 2). By incorporating their conical-wavevector dynamics into holographic synthesis, our framework transfers propagation-invariant behaviour from prescribed analytical beam families to user-defined transverse fields. Because many optical functions are defined by their transverse structure—including images, colour patterns, polarization textures, trapping landscapes and processing profiles—this transfer allows task-specific structures to acquire propagation robustness without being replaced by prescribed analytical beams, with potential relevance to bioimaging[19,20], precision laser processing[21,22], optical manipulation[17,18], nonlinear optics and atom optics[10,11]. In the demonstrated reconstructions, this flexibility is accompanied by reduced characteristic sidelobe disturbance (Supplementary Note 7). The metasurface demonstrations further show that the same principle extends from monochromatic scalar to polychromatic and vectorial light.

Several practical trade-offs define directions for further development. Narrowing the annular spectrum extends the propagation-invariant range, but also reduces the available spatial-frequency bandwidth and can limit information capacity, reconstruction fidelity and efficiency. Multi-cone spectral allocation improves spatial-frequency utilization by distributing encoded information across multiple annular bands while retaining the designed axial response. The present metasurface devices exhibit measured efficiencies of approximately 2.7–8.7% (Supplementary Notes 9 and 11). These values reflect the complexity of multidimensional wavelength and polarization multiplexing, including channel-energy allocation, finite meta-atom conversion efficiency, fabrication imperfections and residual interchannel crosstalk[46,66,67,69]. The two metasurface implementations also adopt different complexity–efficiency balances. The full-colour architecture uses wavelength-dependent polarization preparation and polarization-selective detection to reduce interchannel crosstalk and achieves a measured total efficiency of approximately 8.7%, at the cost of additional polarization-management optics. The vectorial architecture instead uses grating multiplexing and does not require the same polarization-selective illumination and detection for field generation, but yields measured efficiencies of approximately 2.7% and 3.3% for the tailored and full-Poincaré vectorial fields, respectively. These trade-offs are specific to the present multidimensional metasurface implementations rather than intrinsic requirements of the propagation-invariance principle. Further improvements may arise from inverse-designed meta-atoms, dispersion engineering, higher-efficiency dielectric platforms, improved fabrication accuracy and optimized wavelength/polarization multiplexing[48,59,61,62]. More broadly, this framework extends holographic field synthesis from transverse-field programmability to joint transverse–axial design by adding axial diffraction management as an additional design freedom. Across monochromatic, polychromatic and vectorial fields, the transverse optical function and its diffraction-driven axial evolution can therefore be co-designed according to the task, providing a general route towards task-configurable optical systems and compact multidimensional photonic platforms.

## Methods

**Diffractive evolution in conventional Fourier holography.** A monochromatic optical field $U(x, y, z)$ propagating along the $z$-axis can be represented by its angular spectrum $S(k_x, k_y)$ in the momentum space as:

$$S(k_x, k_y) = \iiint U(x, y, z) e^{-i(k_x x + k_y y + k_z z)} dxdydz = \mathcal{F}_{xyz}\{U(x, y, z)\} \tag{1}$$

where $(k_x, k_y, k_z)$ denote Cartesian coordinates in the momentum space, and the subscript "$xyz$" of "$\mathcal{F}$" specifies the three dimensions of Fourier transform. The field $U(x,y,z)$ satisfies the Helmholtz Equation under the momentum-space constraint $k_z = \sqrt{k^2 - k_x^2 - k_y^2}$ , excluding evanescent waves. In conventional Fourier holography, a target intensity distribution $I_{target}(x, y)$ is encoded by combining its field with a diffuser phase $\varphi_{diffuser}(x, y)$ , such as a random phase distribution. Substituting $U(x, y, z = 0) = \sqrt{I_{target}(x, y)} \exp(i\varphi_{diffuser}(x, y))$ into Eq. (1) gives the angular spectrum of the Fourier hologram:

$$S(k_x, k_y) = \mathcal{F}_{xy}\{\sqrt{I_{target}(x, y)} \exp(i\varphi_{diffuser}(x, y))\}. \tag{2}$$

The diffuser phase distributes the target-field information across the Fourier domain[23], allowing a phase-only hologram, or kinoform, to be obtained from $\arg(S(k_x, k_y))$. Broad spatial-frequency support enables flexible reconstruction of complex transverse fields near the designed plane. During propagation, however, the constituent plane waves accumulate different axial phases because their longitudinal wavenumbers differ. The resulting relative dephasing governs the diffractive evolution of the reconstructed field and causes a rapid loss of fidelity away from the designed reconstruction plane (Figs. 1A and 1B).

Depth of field provides a quantitative measure of this axial evolution. In conventional Fourier or Fresnel holography, it is determined by the wavelength, propagation distance, hologram pixel parameters and spatial-frequency distribution of the encoded field, and typically lies within the millimetre-to-centimetre range[8,9,34]. This rapidly varying axial response is useful for axial selectivity in stacked or multiplane three-dimensional displays. It also makes the reconstruction sensitive to axial displacement: when the observation plane, such as the eyebox in a near-eye display[3], or the display medium moves beyond this range, the reconstructed field rapidly blurs, fades or loses fidelity[2,4,6].

**Encoding propagation invariance through conical-wavevector design.** To encode a propagation-invariant axial response, the constituent plane-wave components are designed to accumulate a common axial phase factor during propagation. This condition is satisfied when their longitudinal wavenumber is constrained to a prescribed value $k_{z0}$, as in established non-diffracting structured light. The corresponding field can be expressed as:

$$U(x, y, z) = \sqrt{I_{target}(x, y)} \exp(i\varphi_{diffuser}(x, y)) \exp(ik_{z0} z), \tag{3}$$

where $k_{z0}$ is the prescribed longitudinal wavenumber. Substituting Eq. (3) into Eq. (1) gives the corresponding momentum-space angular spectrum:

$$S(k_x,k_y)=\mathcal{F}_{xy}\{\sqrt{I_{target}(x,y)}e^{i\varphi_{diffuser}(x,y)}\}\mathcal{F}_z\{e^{ik_{z0}z}\}$$
$$=\mathcal{F}_{xy}\{\sqrt{I_{target}(x,y)}e^{i\varphi_{diffuser}(x,y)}\}\delta(\sqrt{k^2-k_x^2-k_y^2}-k_{z0}). \tag{4}$$

This spectrum can be understood as a conventional Fourier hologram modulated by an annular constraint in the transverse spatial-frequency domain (Figs. 1B and 1D). The annular spectrum corresponds to a conical distribution on the ***k***-surface (Fig. 1C). Its constituent plane waves share the same longitudinal wavenumber $k_{z0}$ and therefore acquire a common axial phase $k_{z0}z$ during propagation. Their relative interference condition is consequently preserved, allowing the transverse field to remain invariant, apart from a global phase, over the designed axial range[10–12,74,75].

We incorporate this established conical-wavevector condition into computer-generated hologram synthesis, adding axial diffraction control to the transverse-field programmability of holography (Figs. 1C and 1D). The transverse optical function and its axial evolution can therefore be specified within the same synthesis framework, linking holographic programmability with the propagation robustness of non-diffracting structured light (see Supplementary Note 4).

**Bandwidth control of the propagation-invariant range.** An ideal propagation-invariant field would require an infinitely narrow annular spectrum and correspondingly unbounded aperture and energy[10–12,74,75]. Practical implementations instead use a finite annular bandwidth, producing a finite and selectable propagation-invariant range. To describe this condition, we introduce a real-space axial window function, rect[($z$-$b$)/2$a$], into the conical-wavevector-designed field in Eq. (3). This confines the propagation-invariant region to an axial interval of length 2$a$, centered at $b$, with $z\in$ ($b$-$a$, $b$+$a$). In momentum space, the axial window transforms the ideal delta-function annulus in Eq. (4) into a physically realizable annular sinc profile:

$$S(k_x,k_y)=\mathcal{F}_{xy}\{\sqrt{I_{target}(x,y)}e^{i\varphi_{diffuser}(x,y)}\}\mathcal{F}_z\{e^{ik_{z0}z}rect[\frac{z-b}{2a}]\}$$
$$=\mathcal{F}_{xy}\{\sqrt{I_{target}(x,y)}e^{i\varphi_{diffuser}(x,y)}\}sinc(2a(\sqrt{k^2-k_x^2-k_y^2}-k_{z0}))e^{-i\sqrt{k^2-k_x^2-k_y^2}b}. \tag{5}$$

The phase term $\exp(-i\sqrt{k^2-k_x^2-k_y^2}b)$ determines the axial location $b$ of the propagation-invariant region, whereas the spectral width ($\propto 1/2a$) governs its axial extent (2$a$). This formulation continuously connects two propagation regimes. A broader annular bandwidth permits a larger spread of longitudinal wavenumbers and therefore produces faster axial evolution, approaching the short-range response of conventional Fourier holography. Narrowing the bandwidth reduces the longitudinal-wavenumber spread and progressively extends the propagation-invariant range.

The annular bandwidth therefore provides a direct means of managing diffractive evolution and selecting the desired axial response of the reconstructed field. The corresponding DoF can be configured for short-range operation when axial selectivity is required, such as optical sectioning[7,8] and multilayer holographic displays[9], or for long-range operation when propagation robustness and defocus tolerance are required. For experimental implementation, the annular sinc profile is approximated by an annular rectangular function with an equivalent full width at half maximum (see Supplementary Note 5).

**Multi-cone spectral allocation for propagation-invariant multiplexing.** A single narrow annular band extends the propagation-invariant range but confines the encoded information to a limited spatial-frequency region. This reduces the number of available Fourier-domain pixels and can limit transverse information capacity and optical efficiency. To mitigate this trade-off, we generalize the design from one conical band to multiple distinct bands, implemented as concentric annuli in the Fourier domain (Figs. 2I–2L).

Each annulus corresponds to a different longitudinal wavenumber and therefore to a distinct conical shell on the ***k***-surface. Within each band, the constituent plane waves satisfy the designed propagation-invariant condition. Different field components can consequently be assigned to separate conical bands within a single kinoform. This multi-cone spectral allocation improves spatial-frequency utilization, while enabling independent propagation-invariant fields to be multiplexed within one hologram (Fig. 2). The same strategy is used in Figs. 3 and 4 to balance propagation range, reconstruction fidelity, information capacity and efficiency.

**Speckle optimization under annular spectral constraints.** The diffuser phase $\varphi_{diffuser}(x, y)$ in Eq. (3) distributes the target-field information across the Fourier domain[23], allowing annular Fourier holograms, or phase-only kinoforms, to reconstruct user-defined propagation-invariant fields for different annular radii (see Fig. S3). Random diffuser phases, however, can introduce speckle noise (Fig. S4). We mitigate this noise using an optimization algorithm tailored to the annular spectral constraint. The algorithm jointly enforces the prescribed spectral support in the Fourier domain and maximizes reconstruction fidelity across multiple axial planes (see Supplementary Note 6). A high-speed rotating diffuser is introduced during image acquisition to reduce speckle associated with coherent illumination[3].

A quantitative comparison with conventional digital holography under identical conditions is provided in Supplementary Note 7 and Supplementary Movie 4. The optimized annular design increases the propagation-invariant range, quantified by DoF, by approximately 65 times. This improvement results from controlling the longitudinal-wavenumber distribution through the designed annular spectrum. Conventional iterative holographic algorithms can improve transverse reconstruction fidelity, but do not by themselves prescribe the axial propagation response associated with a broad angular spectrum[9]. Here, reconstruction optimization is combined with direct spectral control so that both transverse fidelity and axial propagation behaviour are incorporated during hologram synthesis.

**Quantification of propagation invariance by depth of field.** Depth of field is used here as the experimental measure of propagation invariance[9]. It is defined as the axial range over which the reconstructed field maintains an acceptable similarity to the target field[23]. Specifically, the DoF is estimated as the interval over which the Pearson correlation coefficient (PCC) between the target and reconstructed intensity distributions remains above 90% of its peak value. The PCC between two matrices $X$ and $Y$ is defined as $\mathrm{PCC}(X, Y) = \mathrm{cov}(X, Y) / \sigma_X \sigma_Y$, where $\mathrm{cov}(X, Y)$ is their covariance, and $\sigma_X$ and $\sigma_Y$ are their standard deviations. For polychromatic and vectorial fields, the reported propagation range is determined from channel-resolved intensity and polarimetric measurements, with the most rapidly degrading spectral or polarization component setting the usable range (Supplementary Notes 9, 11 and 12).

**Metasurface fabrication.** The fused-silica wafer was ultrasonically cleaned in acetone, ethanol, and deionized water. Subsequently, a 1000 nm layer of silicon nitride ($Si_3N_4$) was deposited onto the cleaned fused quartz substrate using a plasma-enhanced chemical vapor deposition system (Oxford PlasmaPro 100 PECVD). A conductive, positive-tone electron-beam resist (ZEP520A, Zeon), approximately 200 nm thick, was then spin-coated onto the silicon nitride surface. The metasurface pattern was defined in the resist layer via Electron Beam Lithography (Elionix ELS-F125-G8 EBL system). After exposure and development, an etch mask was formed by electron-beam evaporation for the $Si_3N_4$ film. Etching was performed using an Inductively Coupled Plasma Reactive Ion Etcher (ICP-RIE) with $O_2$ and $CHF_3$ gases. Finally, the residual Cr mask was etched away with a cerium ammonium nitrate solution, completing the metasurface fabrication.

## Acknowledgements

This work is supported by the National Program on Key Basic Research Project of China (2022YFA1404300 (to S.W.)), the National Natural Science Foundation of China (12374307 (to J. D.), 12427808 (to H. W.), 12325411 (to S.W.), 62288101 (to S.W.)), the Jiangsu Provincial Key Research and Development Program (BG2024029 (to S.W.)), Joint Funds of the National Natural Science Foundation of China (U24A20313 (to S.W.)), Fundamental and Interdisciplinary Disciplines Breakthrough Plan of the Ministry of Education of China (JYB2025XDXM106 (to S.W.)). Work done in Hong Kong is supported by RGC Hong Kong (AoE/P-502/20 (to T.L.)). This work is also supported by the Postdoctoral Fellowship Program of CPSF under Grant Number GZC20261862 (to. W.Y.), Jiangsu Funding Program for Excellent Postdoctoral Talent (to. W.Y.),

## Author Contributions

W.Y. T.L. and J.D. conceived the original idea and concept. W.Y. carried out the calculations and simulations. T.L. designed, fabricated and characterized the samples. W.Y. built the experimental system and performed the experiments. Z.W. and Y.Z. assisted in the experiments. H.W., S.W. and J.D. supervised the project. All authors contributed to the discussion and writing of the manuscript.

## Competing Interests

The authors declare no competing interests.